\documentstyle[11pt,aaspp4,psfig]{article}

\begin{document}

\title{Observational Constraints on Microwave Anisotropy from Point Sources}

\author{Eric Gawiser\footnote{gawiser@astron.berkeley.edu}, 
Andrew Jaffe, \& Joseph Silk}
\affil{
Departments of Physics and Astronomy and Center for Particle 
Astrophysics, 
   University of California, Berkeley, CA 94720}
\authoremail{gawiser@astron.berkeley.edu}






\begin{abstract}

Applying 
basic physical
principles to recent observational results, 
we derive upper and lower  
limits on microwave anisotropy from point sources over the range of 
frequencies 10-1000 GHz.  We examine the level of noise 
in the observations as a possible indication of source confusion 
at subarcminute scales.  
We also derive an upper limit on microwave anisotropy caused by 
the sources
responsible for the Far-Infrared Background radiation detected in FIRAS 
data.  
Our upper limit on point source confusion of
$\Delta T / T = 10^{-5}$ for a 10$'$ beam at 100 GHz would 
cause severe foreground contamination for CMB anisotropy observations, 
although the actual contamination level is probably much lower.  
This upper limit constrains 
the long-feared possibility of an undetected 
population of sources with emission peaking near 100 GHz.  
Source detections closer to 100 GHz are
needed to improve our knowledge of galaxy evolution at high redshift
and to predict the level of point source confusion.

\end{abstract}


\section{Introduction}

	The detection of anisotropy in the Cosmic Microwave Background
by COBE DMR 
(Smoot et al. 1992) and several other instruments has generated
interest in measuring CMB anisotropy on all angular scales with the goal
of determining cosmological parameters.  
Improved instrumentation
and the upcoming MAP (Microwave Anisotropy Probe) and Planck 
Surveyor\footnote{http://map.gsfc.nasa.gov and http://astro.estec.esa.nl/Planck/}
satellite missions focus current attention on 
angular scales between one-half and one-tenth of a degree, and there is 
theoretical motivation for undertaking future observations 
at even smaller scales (Hu \& White 1997, Metcalf \& Silk 1998, 
Jaffe \& Kamionkowski 1998).  

Because the antenna temperature 
contribution of a point source is inversely proportional to  
the solid angle of the beam,
observations at higher 
angular resolution 
are more sensitive to extragalactic foregrounds, including 
radio sources, low- and high-redshift 
infrared-bright galaxies, and the Sunyaev-Zel'dovich 
effect from galaxy clusters.  
The dominant contribution of the Galaxy to microwave 
anisotropy is from diffuse emission (Toffolatti et al. 1998,
Finkbeiner et al. 1998).   
Until recent SCUBA observations, almost all sources observed from 
10-1000 GHz were selected at higher or lower frequencies, so there 
was little direct knowledge of point source populations with emission 
peaking in this wide frequency range.  
Blain, Ivison, and Smail (1998, see also Blain, Ivison, Smail, \& Kneib 1998
 and Scott \& White 1998) use models for high-redshift galaxies
normalized to SCUBA counts at 353 GHz to predict anisotropy 
from extragalactic point sources 
down to 100 GHz, but this extrapolation is 
model-dependent.  Previous predictions of the total point source
contribution (Toffolatti et al. 1998, 
Toffolatti et al. 1995, Franceschini et al. 1989, Wang 1991)
 used galactic evolution models with specific 
assumptions about dust
temperatures and luminosity evolution 
to predict the level of extragalactic foreground.  More
phenomenological approaches 
(Gawiser \& Smoot 1997, hereafter GS97, 
Sokasian et al. 1998, hereafter SGMS,
and Tegmark \& Efstathiou 1996) 
lack information on infrared galaxies at high redshift and on dim but 
numerous radio sources.    
 
	Cosmic microwave background observations contain contributions
to anisotropy from two groups of point sources.  The bright sources
at a level of at least $ 5 \sigma$ 
($\sigma$ is the quadrature sum of instrument
noise, CMB fluctuations, diffuse Galactic emission,
 and underlying point source
fluctuations) can be detected individually and eliminated by 
masking the pixels containing them.  This detection limit can 
be lowered by using prior 
knowledge of the locations of bright sources obtained from 
extrapolating far-infrared and radio frequency observations as
described in GS97 and SGMS (as well as filtering, fourier transform, 
 and wavelet techniques; 
see Tegmark \& de Oliveira-Costa 1998, Ferreira \& Maguiejo 1997, 
Tenorio et al. 1998).  Numerous dimmer sources 
will add to anisotropy but cannot
be detected without performing further observations at higher resolution 
at nearby frequencies.
For most planned CMB observations, these simultaneous observations will be 
difficult due to large sky coverage at high resolution of
the primary instrument (although Planck's wide frequency coverage 
will help with foreground subtraction.)

We utilize recent sub-arcminute resolution
observations to constrain the contribution to anisotropy from this
second group of point sources that will inevitably contaminate
measurements of CMB anisotropy.    
Recent observations using BIMA by Wilner \& 
Wright (1997) detected no sources brighter than 3.5 mJy in the 
15 arcmin$^2$ of the Hubble Deep Field (HDF) at
4.7$''$ resolution at 107 GHz.  We combine this constraint with the 
counts of sources detected in blank fields at 353 GHz by SCUBA (Hughes et al. 
1998, Barger et al. 1998, Eales et al. 1998)
and at 8.4 GHz with the VLA 
(Richards et al. 1998, Fomalont et al. 1997), 
with blank field upper limits from BIMA/OVRO 
at 28.5 GHz (Holzapfel 1998, Carlstrom 1998), 
SuZIE at 142 GHz (Church et al. 1998), 
IRAM at 250 GHz (Kreysa 1998, 
Grewing 1997), SCUBA at 667 GHz (Hughes et al. 1998),  
and CSO at 850 GHz (Phillips 1998), and with the 
detection of Far-Infrared Background radiation 
in FIRAS data (Puget et al. 1996,
Burigana \& Popa 1997, Fixsen et al. 1998).  

It has long been feared that a population of 
sources with spectra peaking near 100 GHz, due to self-absorbed radio 
emission or thermal emission at very high redshift, might remain
undetected by radio and far-infrared observations while contributing 
significantly to measurements of CMB
anisotropy.  
Now that high-resolution observations are available in the 
frequency range relevant to CMB anisotropy observation, we
set upper and lower limits on point source confusion between 
10 and 1000 GHz 
by assuming 
that the emission of point sources originates from 
synchrotron, free-free, thermal dust, and spinning dust grain emission.  

\section{Extragalactic Point Sources}	

The main emission mechanism 
of bright far-infrared 
 sources is graybody reradiation of starlight and/or Active Galactic Nuclei 
(AGN) radiation absorbed by dust. 
GS97 predict the 
level of microwave anisotropy from the 5319 low-redshift
infrared-bright galaxies in the IRAS 1.2 Jy survey (Fisher et al. 1995). 
We expect there to be numerous higher-redshift starburst 
galaxies like the prototypes Arp 220, F 10214+4724, SMM 02399-1236
(Ivison et al. 1998), and APM 08279+5255 (Lewis et al. 1998) 
which generate
similar dust emission, and with their spectra redshifted considerably 
these sources could easily be missed by far-infrared surveys and yet 
make significant contributions to the microwave sky.  There may  
exist a population of ultraluminous proto-elliptical galaxies which 
cannot be described by models using smooth evolution of the IRAS 
luminosity function.  Recent detections of the 
Far-Infrared Background radiation 
and of submillimeter sources by SCUBA (Smail, Ivison, \& Blain 1998; Smail, 
Ivison, Blain, \& Kneib 1998)
 give us the first clues about
the nature and abundance of these high-redshift objects.

	A separate population of extragalactic point sources 
are radio-loud, typically elliptical galaxies or AGN.    
Radio sources which have nearly flat spectra up through microwave 
frequencies are called blazars, a class which includes 
radio-loud 
quasars and BL Lacertae objects where synchrotron self-absorption due 
to the opacity of the dense nuclear regions at low frequencies
 prevents the spectrum 
from falling with frequency.  SGMS examine 2200 bright
radio sources in detail, but there are over ten thousand 
of these sources which are bright enough to have an impact on 
arcminute-resolution microwave observations.  

For instruments of resolution $\geq 10'$, galaxy clusters will be  
unresolved and will provide an additional family of point sources 
via the Sunyaev-Zel'dovich
 effect (Sunyaev \& Zel'dovich 1972).   
The observations used here are basically insensitive to SZ clusters 
as the fields have been chosen to avoid known clusters and are 
typically observed at sub-arcminute resolution. 
Anisotropy from SZ sources is not expected to seriously impair CMB anisotropy 
observations (Refregier, Spergel, \& Herbig 1998, Aghanim et al. 1997).

\section{Analysis}

We assume for these calculations that observations use pixels
of width equal to the FWHM of their beam.  Overpixelization will 
lead to a small correction in the level of anisotropy 
and makes it easier to distinguish point sources,
which contribute to several pixels, from instrument noise, which is often 
uncorrelated between neighboring pixels.  

We can rigorously predict the fluctuations due to
sources using the techniques of $P(D)$ analysis (Scheuer 1957,1974; 
Condon 1974; Franceschini et al. 1989; Toffolatti et al. 1998). To
begin, we must estimate the cumulative flux distribution of sources, 
$N(>S)=\int_S^\infty N(S)\; dS$.  
The SCUBA results give a list of sources and their
fluxes with error bars; they also provide a limit on the
low-flux tail of the distribution from their measured residual
fluctuations (Hughes et al. 1998). 
We estimate $N(>S)$ directly, using a
Gaussian of width given by the error on the observed flux for each
source.  
We calculate $2\sigma$ error bars on $N(>S)$ 
for this estimated distribution by having the fluctuations in number be 
consistent with Poissonian fluctuations for each cumulative distribution.
We use a top-hat experimental beam to 
convert $N(>S)$ to an observed flux distribution. We then convert
this to the probability distribution, $P(D)$, of getting a total flux,
$D$, in the beam using the elegant formulae of Scheuer (1957; 1974) and
Condon (1974).  Whereas the integrated background is determined by 
the slope and low-flux cutoff of $N(>S)$, the anisotropy is dominated 
by the brightest sources seen by SCUBA.  

We consider both the detected sources and the rms noise in 
the instrument.  The observed instrument noise is usually roughly 
Gaussian with mean near zero.  This provides a good upper limit on 
confusion from undetected sources 
because one can bury only about half that noise in anisotropy from 
dim sources without increasing the mean noise level by much or making the 
noise distribution noticeably non-Gaussian.
Hughes et al. report a noise level of 0.45 mJy
per 8.5$''$ beam. To allow for the possibility that a large
fraction of this may actually be from sources, we define the total flux,
$y$, the sum of $D$ and this noise contribution.  We take
the noise distribution to
be a zero-mean Gaussian with the reported variance, scaled by 
the desired beam area.  The distribution of $y$
is just the convolution of $P(D)$ and the noise distribution.  
From $P(D)$ or $P(y)$ we determine the impact on CMB measurements by
estimating the variance 
($\sigma_D$ and $\sigma_y$).  Our $2\sigma$ upper and lower limits from 
SCUBA at 353 GHz are 
67 mJy and 8 mJy respectively.  
Such careful calculations are not strictly necessary; 
the following easily-reproduced back-of-the envelope
calculation is accurate to within 
a factor of two, adequate for present purposes.

Our upper and lower limits correspond to $2\sigma$ 
 confidence levels from the reported observations.  
If an observational field contains 
$N_{obs}$ sources, we estimate the upper/lower 
limit 
on the number of sources $N$ in a typical such field on the sky using 
$N_{obs} = N \pm 2 \sqrt{N}$, which leads to limits on the fluctuation of the 
number of sources in a typical field on the sky of $\sqrt{N} = 
\sqrt{N_{obs}+1} \pm 1$.  
 
For $N$ sources with flux $S$ per beam, the rms flux anisotropy on the sky 
is 

\begin{equation}
 \Delta S = S \sqrt{N} \; \;  
\end{equation}

\noindent
in the Poisonnian limit of large N.  Toffolatti et al. (1998) 
predict a negligible contribution from
non-Poissonian clustering of sources for beams of 10$'$ and larger.  Scott 
\& White (1998), however, suggest that clustering will lead to fluctuations 
twice as large as Poissonian fluctuations for a 10$'$ beam, with less 
enhancement at higher resolution.  
For N$<1$ (one source per several
beams), we have only a few pixels receiving flux, the mean flux is $NS$, and

\begin{equation}
  \Delta S = \sqrt { N (S-NS)^2 + (1-N)(NS)^2 } = S \sqrt{ N - N^2} \; \; , 
\end{equation}

\noindent
which also tends towards $S \sqrt{N}$ 
as N becomes small.  

We extrapolate our upper and lower limits 
from an observed frequency by using the 
most extreme known physical emission mechanisms in that frequency 
range; the fastest the flux should fall is as 
very steep spectrum synchrotron emission, i.e. $\nu ^{-2}$ (Steppe 
et al. 1995), or above 300 GHz as a Wien tail with $\nu^1$ emissivity,
i.e. 
$\nu^3 / (\exp(h \nu / k T_{CMB}) - 1)$, 
since it is unreasonable 
for a cosmological object to have an effective temperature less than 
T$_{CMB}$.  
Conversely, the fastest a spectrum should be able to rise is as 
Rayleigh-Jeans thermal emission 
with $\nu^2$ emissivity, i.e. as $\nu^4$.   Free-free and spinning dust grain
emission (Draine \& Lazarian 1998) produce less conservative extrapolations.

The way to maximize anisotropy for the observed integrated 
Far-Infrared Background
is to make individual sources as bright as possible. 
The low emissivity ($\nu^{0.6}$) fit by Fixsen et al. (1998) means that 
no one graybody spectrum (emissivity between $\nu^1$ and $\nu^2$) can be 
responsible for the FIRB.  Therefore, we set upper limits on the 
anisotropy at a given frequency by making hypothetical sources whose spectra
peak at that frequency be as bright as possible.  
The brightness of these high-z IR sources 
is constrained by requiring their dust to have temperature 
greater than 20K (since low-z inactive spirals have 20K dust) and 
greater than 3K(1+z) (so that the dust is never 
colder than the CMB at that redshift), and to have a  
bolometric luminosity 
no greater than that of a quasar ($10^{39}$ W).  We also examine 
a second model where the luminosity constraint is raised to $10^{41}$W, the 
likely luminosity of APM 08279+5255 once lensing is accounted for (Lewis 
et al. 1998).  
Using these constraints, we predict an upper limit of 
$\Delta T / T = 10^{-6} (10^{-5})$ 
for a 10$'$ beam at 200 GHz for the high-z IR population of 
luminosity $10^{39}$W ($10^{41}$W)
 whose total emission generates the FIRB.  However, this upper limit 
is less robust than those from direct observations, because there 
could be separate source populations, one which yields the integrated
background but small fluctuations on the relevant angular scales, 
and another which dominates the flux
anisotropy but produces only a small fraction of the FIRB.

Flux variation, measured in Jy ($1$ Jy $= 10^{-26} W/m^2/$Hz), 
is converted to antenna temperature $T_A$ by 

\begin{equation}
 T_A = S \frac{\lambda^2}{2 k_B \Omega}\; \;,  
\end{equation}

\noindent 
where $k_B$ is Boltzmann's constant, $\lambda$ is the wavelength, and 
$\Omega$
is the effective beam size.  

Small fluctuations in antenna 
temperature can be converted to effective thermodynamic 
temperature fluctuations about a mean temperature $T_{CMB}$ using

\begin{equation}
 \frac{dT_A}{dT} = \frac { x^2 e^x} {(e^x - 1)^2} \; \; ,
\end{equation}

\noindent
defining $x \equiv h \nu / k T_{CMB}$.  
This yields an equivalent thermodynamic temperature variation which
 scales as fwhm$^{-1}$ for a given flux anisotropy on $10'$ scales:

\begin{equation}
\frac{\Delta T}{T_{CMB}} = \Delta S_{10'}(Jy) 
\left ( \frac{fwhm}{10'} \right )^{-1} (5\times10^{-4})
 \left ( \frac {(e^x - 1 )^2}{x^4e^x} \right ) \;  \;  .
\end{equation}

\section{Results}

Table 1 shows our upper limits for the possibility 
of dim sources buried in the instrument noise of non-detections, and 
Table 2 lists source detections and the resulting limits on 
$\Delta S_{10'}$.
Figure 1 shows our upper and lower limits for
flux anisotropy 
from point sources from 10-1000 GHz, as well 
as the results for the extreme models of the Far Infrared Background
radiation.

We plot
the resulting limits on temperature anisotropy in Figure 2 for a
range of angular scales and frequencies.  The less robust nature of 
the FIRB constraint 
prevents us from using this as an upper limit in Figure 2.  
Because the angular power spectrum $C_\ell$ 
of 
Poissonian distributed point
sources increases with multipole $\ell$ 
relative to the expected CMB angular power spectrum (Scott \& White 1998),
an rms $\Delta T / T$ from point sources close to $10^{-5}$
will seriously impair
the measurement of the CMB angular
power spectrum on the smallest angular scales, whereas a value less than 
$10^{-6}$ means that foreground contamination is not a major concern.
The 10$'$ lower limit shows that $\Delta T / T < 10^{-6}$ is only possible 
from 20-300 GHz, and the upper limit for 10$'$ shows that $\Delta T / T < 
10^{-6}$ at 30 GHz and $\Delta T / T \simeq 10^{-6}$ at 250 GHz.  The limits
are much less stringent near 100 GHz, where a pathological population of 
point sources could lead to anisotropy up to $10^{-5}$.  Typical radio 
and far-IR sources that fall within these limits at 30 and 250 GHz will 
end up much closer to the lower limit near 100 GHz, however.  
Our upper limit constrains all types of point sources, including any 
hypothetical high-latitude or halo Galactic point sources.  
Our limits diverge considerably near 100 GHz, so while they 
are compatible with the model-dependent extrapolations of 
Blain et al. and Toffolatti et al.,   
they would also be compatible with significantly different extrapolations.
Scott \& White (1998) indicate that clustering 
may lead to a factor of two amplification of our predictions for a 10$'$ beam; 
the correction is less for higher resolution.  

The blank fields observed by VLA, BIMA, IRAM, SCUBA, and CSO  
were chosen 
to avoid known bright point sources.  
Therefore, for observations which avoid known bright sources or 
mask the pixels containing them, Figure 2 
gives full upper and lower 
limits on point source anisotropy.  
GS97 and SGMS analyze the contribution of 
bright IR and radio point sources, 
respectively, 
so for a randomly chosen location on the sky 
the expected anisotropy is the quadrature sum of the 
anisotropies from those types of bright sources and our result
 in Figure 2.  
Figure 3 adds in results for known bright
sources from GS97 and SGMS for a 10$'$ beam and 
shows the results for MAP and Planck 
after subtracting sources detected at $5 \sigma$. 
Since 
$ \Delta T / T$ is strongly influenced
by a few bright pixels due to the highly non-Gaussian distribution, 
the values are significantly lower after bright source subtraction.    
Figure 3 shows that for a 10$'$ beam without source subtraction, the 
point source anisotropy will be $\geq 10^{-6}$ at all frequencies.  
From 70-200 GHz, the 
upper limit from Figure 2 dominates the anisotropy from known bright 
radio and IRAS sources.  MAP and Planck can detect the brightest few 
hundred sources at each frequency (see SGMS) so the upper and lower limits 
for the satellites diverge over a wider frequency range, making the impact 
of our uncertainty about the level of anisotropy from dim but numerous 
point sources a significant problem in predicting foreground contamination.  
The highest-frequency Planck channels can detect nearly all 5319 IRAS 1.2 Jy 
sources, so it is the dimmer 
high-redshift IR galaxies constrained by SCUBA that 
dominate their source confusion.  
Our limits here treat each channel independently, but it will be possible
to detect bright sources at particular frequencies and mask the corresponding
pixels in all channels.  This will enhance the importance of dim but 
numerous sources relative to known bright sources but will reduce the 
overall level of foreground contamination.

\section{Discussion}

We find impressive agreement between the SCUBA observations, the IRAM and 
SCUBA upper limits, and the upper limit for flux anisotropy produced by 
$10^{39}$W sources which generate the integrated Far-Infrared Background 
shown in Figure 1.  
This is  
consistent with the FIRB being produced by the SCUBA sources (since the upper 
limits and the detections differ by only a factor of two), and this indicates 
that the FIRB sources must be close to maximizing their anisotropy i.e. they 
are highly luminous but not too numerous.  
The $10^{41}$W model, however, 
predicts more anisotropy than is consistent with the observed SCUBA 
source counts and IRAM and SCUBA upper limits, suggesting that starburst 
galaxies 
like APM 08279+5255 are more luminous than 
typical FIRB sources.  This conclusion is also 
supported by the near-blackbody spectrum of APM 08279+5255 in the 
sub-millimeter; there is no way to sum such spectra at various redshifts 
and produce a graybody of emissivity 0.6 as is seen for the FIRB (Fixsen 
et al. 1998).  
The IRAM upper limit is low enough 
to show that the far-IR sources detected by SCUBA have rising spectra, so 
this is further evidence that their emission is thermal in origin.  

The CMB anisotropy damping tail on arcminute scales is a sensitive probe 
of cosmological parameters and has the potential to 
break degeneracies between models which explain the larger-scale anisotropies
(Hu \& White, Metcalf \& Silk).  The expected level of temperature 
anisotropy 
is $\Delta T / T \simeq 10^{-6}$, 
which Figure 2 indicates may be enough to 
dominate the point source confusion from 30-200 GHz.  The upper limit on 
point source confusion, however, would completely swamp the fluctuations 
of the damping tail, so more knowledge of dim sources is needed before 
we can expect such observations to be feasible.   A high resolution instrument
could use its highest resolution for point source detection and 
subtraction.

The greatest promise for seeing CMB anisotropies through the 
obscuration of point source confusion occurs near 100 GHz, but 
this is also the frequency range where we know the least about 
the true level of foreground anisotropy on the sky.  
Our upper limits for 10$'$ near 100 GHz give us confidence that useful 
information will be obtained from CMB anisotropy 
observations, but it remains possible that point sources will cause 
thermodynamic fluctuations 
roughly equal to 
the intrinsic CMB fluctuations.  Since the point source 
fluctuations come from the highest multipoles, this could seriously impair
attempts to measure cosmological parameters from the 
CMB angular power spectrum.
Thus, 
further high-resolution observations 
of blank fields at frequencies near 100 GHz are critical in order 
to determine the actual level of point source confusion,
and CMB anisotropy 
analysis methods must account carefully for contamination from 
point sources.

\section{Acknowledgments}

We thank Andrew Blain for his comments on this paper, and 
Paola Platania, Davide Maino, Gianfranco De Zotti, Carlo Burigana, 
Malcolm Bremer, and Marco Bersanelli for helpful conversations.  
E.G. acknowledges the support of an NSF Graduate Fellowship.

\newpage

\begin{table}[th] \caption{Noise levels in 
high-resolution microwave observations.
  We list the frequency, resolution, noise per 
beam, and the upper limit 
for $\Delta S_{10'}$ that results from assuming that half of this
noise is really produced by unresolved point sources.}
\begin{center}
\begin{tabular}{|r|c|c|c|c|}
\hline
Instrument &  $\nu$ (GHz) & FWHM  & 
   Noise/beam & $\Delta S^{upper}_{10'}$ \\

\hline

VLA	& 8.4	&6$''$ 	&0.0028 mJy &0.14 mJy  \\

BIMA	& 28.5 	&90$''$ 	& 0.12 mJy &0.4 mJy  	\\

BIMA	& 107	&4.7$''$	&0.7 mJy &45 mJy  	\\

SuZIE   & 142   & 100$''$       & 10 mJy & 30 mJy       \\

IRAM	& 250	&11$''$ 	&0.5 mJy &14 mJy  	\\

SCUBA	& 353	&8.5$''$ 	&0.45 mJy &16 mJy 	\\

SCUBA	& 667	&7.5$''$ 	&7 mJy	&280 mJy 	\\
	
CSO	& 857  	&10$''$ 	&100 mJy &3000 mJy  	\\

\hline

\end{tabular}
\end{center}
\end{table}


\nopagebreak


\begin{table}[bh] \caption{Microwave source detections.
Upper and lower limits correspond 
to the observed fields being 2 $\sigma$ Poissonian fluctuations above or below
the typical source density on the sky (see text).  
The totals include the noise 
totals given in Table 1 added in quadrature with the limits from each source 
population.  The range of sources in the $>3$mJy SCUBA bin allows for 
the incompleteness correction suggested by Eales et al. and the approximate 
number in the $1-2$mJy bin is based on the P(D) analysis of Hughes et al.}
\begin{center}
\begin{tabular}{|r|c|c|c|c|c|c|}
\hline
Instrument &  $\nu$ (GHz) & Field size & 
S$_{source}$ & N$_{sources}$ & $\Delta S^{upper}_{10'}$ & 
$\Delta S^{lower}_{10'}$ \\

\hline

VLA	& 8.4	&40 sq. $'$ &$>0.5$ mJy  &3 &2.6mJy &0.9 mJy \\

	& 	& 	&0.05-0.5 mJy & 8 &0.4 mJy  &0.2 mJy \\ 

	& 	& 	&0.009-0.05 mJy & 18 	& 0.1 mJy & 0.08 mJy\\

	& 	& 	&0.006-0.009 mJy & 19 & 0.04 mJy & 0.03 mJy\\

	& 	& 	& TOTAL 	& 	& 2.7 mJy & 0.9 mJy \\

SCUBA	& 353	&46 sq. $'$ & $>3$ mJy & 15-20 &29 mJy &16 mJy	\\

	& 	&9 sq. $'$  & 2-3 mJy & 2    & 12 mJy & 7 mJy   \\

	&       &9 sq. $'$  & 1-2 mJy & $\simeq 18$  & 18 mJy & 15 mJy \\ 

	& 	& 	  & TOTAL   & 		& 40 mJy  & 23 mJy \\

\hline

\end{tabular}
\end{center}
\end{table}

\newpage
\begin{figure}
\figurenum{1}
\centerline{\psfig{file=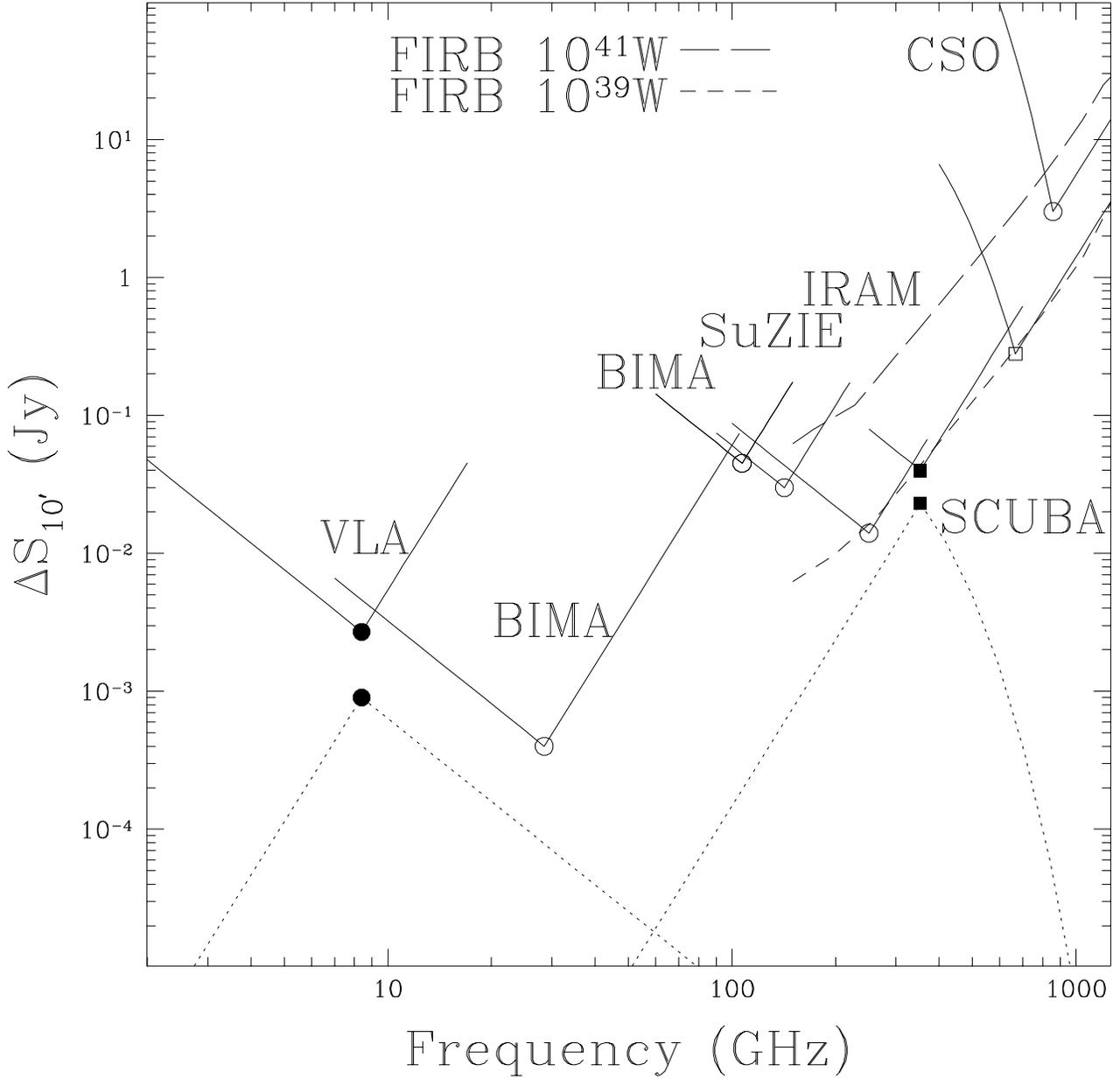,width=7in}} 
\caption{ 
Upper (solid) and lower (dotted) 
limits on flux anisotropy (in Jy) for a 10$'$ beam  
from VLA, BIMA/OVRO, BIMA, SuZIE, IRAM, SCUBA (squares), and CSO.  
Filled points indicate detections, open points are non-detections, 
and the extrapolations are based on steep-spectrum radio emission, 
Rayleigh-Jeans thermal emission with $\nu^2$ emissivity, and 
Wien tail thermal emission with $\nu^1$ emissivity.  The long (short) 
dashed lines
indicate the upper limit on flux anisotropy derived from extreme 
models of the  
Far-Infrared Background radiation with a maximum 
source luminosity of $10^{41}$W ($10^{39}$W).  
}
\end{figure}

\newpage

\begin{figure}
\figurenum{2}
\centerline{\psfig{file=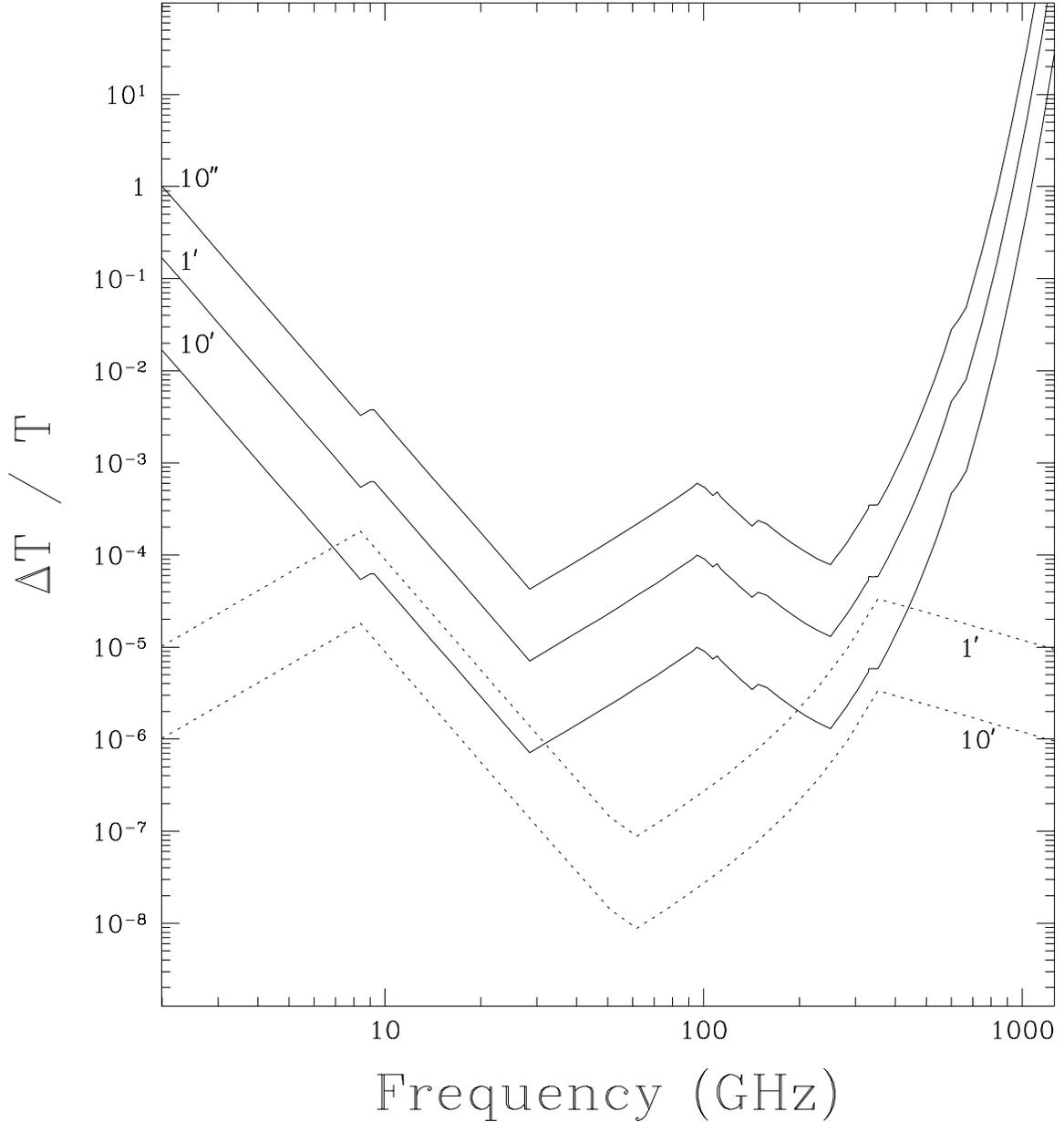,width=7in}}
\caption{
Net upper and lower limits on 
$\Delta T / T$ for $10'$, $1'$, and $10''$ based on the observations
and extrapolated limits shown in Figure 1.  
The lower limit for $10''$ is zero because all sources detected by 
SCUBA and the VLA should also be detected and subtracted 
by future observations at that 
resolution.  The upper limits are based on assuming that the 
combination of instrument noise and CMB fluctuations 
is too high to subtract any of the SCUBA or VLA sources.
  }
\end{figure}
              
\begin{figure}
\figurenum{3}
\centerline{\psfig{file=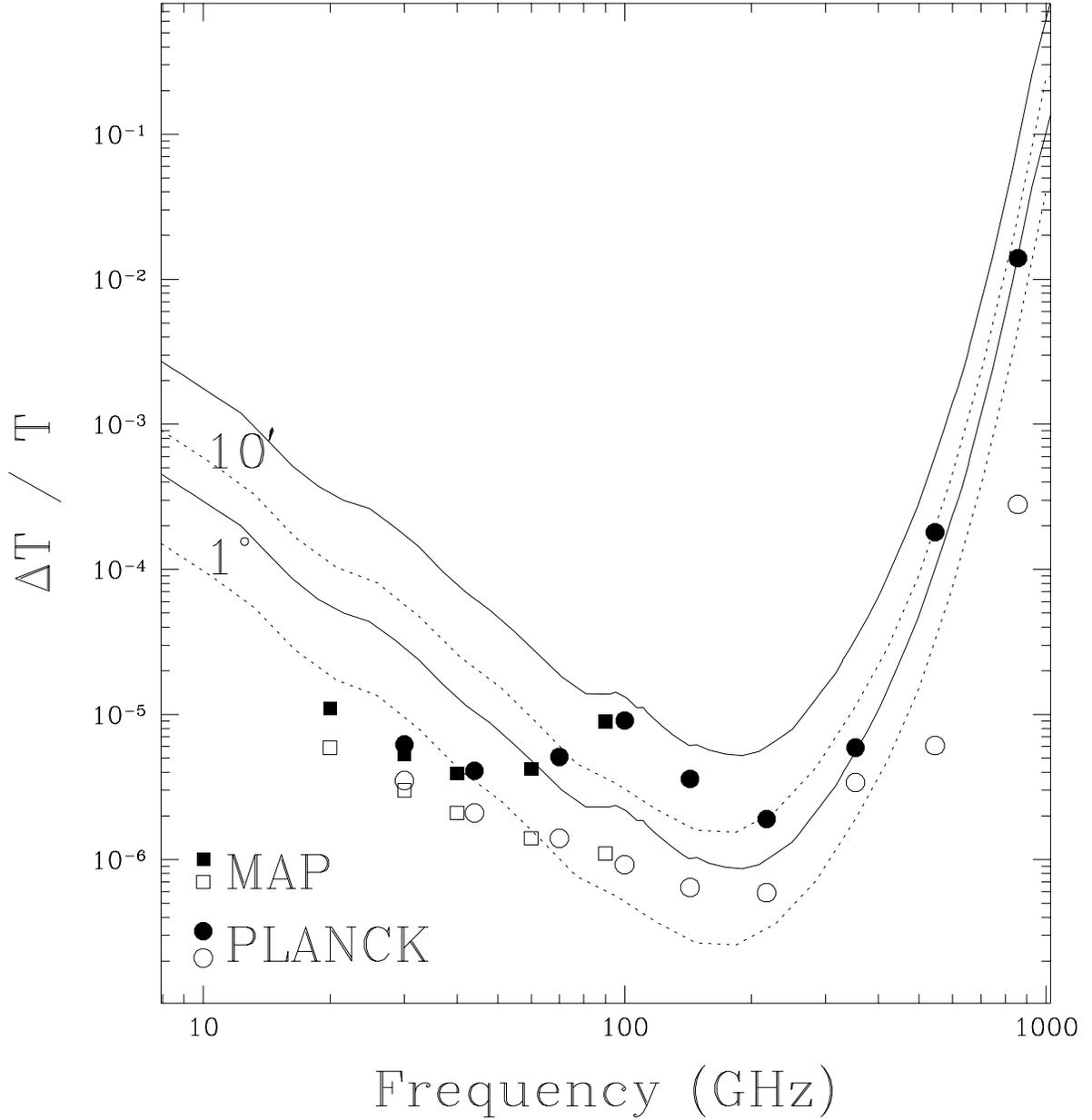,width=7in}}
\caption{
Net upper (solid line) and lower (dotted line) limits for $10'$ and $1^\circ$,
 including anisotropy for known bright sources from GS97 and SGMS with a 
factor of three uncertainty shown.     
5 $\sigma$ subtracted upper (solid) and lower (open) 
limits for MAP (squares) and Planck (circles) are also 
shown.  The channels are treated independently here, although in practice 
they could be combined to produce somewhat lower anisotropy levels.  The 
combination of pixelization and convolving effects discussed in the 
text leads to 1/fwhm scaling for all of these point source populations.    
}
\end{figure}




\end{document}